\documentclass[12pt]{article}
\usepackage{graphicx}

\newcommand\pubnumber{SNSN-323-63}
\newcommand\pubdate{\today}

\def\napoli{Kellogg Radiation Laboratory\\
California Institute of Technology, Pasadena, CA, USA}
\def\support{\footnote{Work supported by the National Science Foundation Grant 1506459}}

\def\Title#1{\begin{center} {\Large #1 } \end{center}}
\def\Author#1{\begin{center}{ \sc #1} \end{center}}
\def\Address#1{\begin{center}{ \it #1} \end{center}}

\newcommand\pubblock{\rightline{\begin{tabular}{l} \pubnumber\\
         \pubdate  \end{tabular}}}
\newenvironment{Abstract}{\begin{quotation}  }{\end{quotation}}
\newenvironment{Presented}{\begin{quotation} \begin{center} 
             PRESENTED AT\end{center}\bigskip 
      \begin{center}\begin{large}}{\end{large}\end{center} \end{quotation}}

\def\beq{\begin{equation}}
\def\eeq#1{\label{#1}\end{equation}}
\def\eeqn{\end{equation}}

\def\beqa{\begin{eqnarray}}
\def\eeqa#1{\label{#1}\end{eqnarray}}
\def\eeqan{\end{eqnarray}}

\let\bar=\overbar

\def\Dslash{\not{\hbox{\kern-4pt $D$}}}
\def\dslash{\not{\hbox{\kern-2pt $\del$}}}

\def\msb{{\bar{\ssstyle M \kern -1pt S}}}

\begin{document}
\begin{titlepage}
\pubblock

\vfill
\Title{Worldwide Search for the Neutron EDM}
\vfill
\Author{ B. W. Filippone\support}
\Address{\napoli}
\vfill
\begin{Abstract}

Existing limits on the electric dipole moment (EDM) of the free neutron have provided critical constraints on new sources of CP violation for more than 60 years. A new round of searches are actively underway with the goal of improving the sensitivity to CP violation by up to two orders-of-magnitude. The status of these new searches will be discussed, including recent progress on the nEDM experiment to be carried out at the Fundamental Neutron Physics Beamline at the Oak Ridge National Laboratory's Spallation Neutron Source.
\end{Abstract}
\vfill
\begin{Presented}
Thirteenth Conference on the Intersections of Particle and
Nuclear Physics (CIPANP2018)
Palm Springs, California, USA, May 29-June 3, 2018
Report No. CIPANP2018-Filippone
arXiv:XXX\\
\end{Presented}
\vfill
\end{titlepage}
\def\thefootnote{\fnsymbol{footnote}}
\setcounter{footnote}{0}

\section{Introduction}

The search for the electric dipole moment (EDM) of the free neutron began more than 65 years ago and yielded an upper limit~\cite{SPR57} of $4\times 10^{-20}$ e-cm at the 90\% confidence level. Because a permanent neutron EDM would violate both parity and time-reversal symmetries (and correspondingly CP violation via the CPT theorem), the search for a neutron EDM continues to the present, with the latest best published upper limit~\cite{NEDM15} of $3\times 10^{-26}$ e-cm at the 90\% confidence level. Future measurements remain highly motivated because their sensitivity to new physics can be dramatic. This results from the very small predicted neutron EDM from the present electroweak Standard Model~\cite{KZ82} and the potential access to CP-violating exchanges of massive virtual particles reaching to mass scales of tens of TeV~\cite{PR05} and in some models up to PeV~\cite{MPR13,AIN15}. In addition, CP violation beyond that present in the Cabbibo-Kobayashi-Maskawa quark weak mixing matrix, could also provide a resolution to the puzzle of the matter-antimatter asymmetry of the universe as suggested many years ago by Sakharov~\cite{S67}. 

A new round of searches  for the neutron EDM are underway worldwide with the goal of improving the sensitivity to CP violation by up to two orders-of-magnitude. These new experiments use a range of promising technologies to improve the experimental sensitivity. We will discuss the technical advances and outline the status of these new searches, including recent progress on the nEDM experiment to be carried out at the Fundamental Neutron Physics Beamline at the Oak Ridge National Laboratory's Spallation Neutron Source (SNS). 

\section{Technologies for Neutron EDM Measurements}

We can characterize the sensitivity to a neutron EDM in terms of the 1-$\sigma$ uncertainty - $\sigma_d$ - in the measured dipole moment $d$ 

$$
\sigma_d \propto \frac{\hbar}{|\vec{E}|T_m\sqrt{N}}
$$
where  $|\vec{E}|$ is the magnitude of the applied electric field, $T_m$ is the observation time for a single measurement and 
$\sqrt{N}$ is the statistical shot noise in terms of the number of observed/detected particles $N$. The proportionality constant depends on the signal-to-noise and details of how the measurement is performed. The dependence on $T_m$ results because the measurement of an EDM is generally a measurement of the E-field-dependent precession frequency of neutrons that are also precessing in an applied magnetic field. Thus improvements in sensitivity require 
increases in one or all of these parameters while also minimizing the potential systematic uncertainties. Some of major systematic uncertainties include leakage currents associated with the electric field, other variations in the ambient magnetic field and the false EDM associated with gradients in the magnetic field; sometimes called the geometric phase effect~\cite{GEOALL}. To increase EDM sensitivity these systematic effects can be reduced through the use of magnetometers, with negligible EDM themselves, that monitor the ambient B-field. Generally atomic magnetometers such as $^{199}$Hg,$^{133}$Cs, $^{129}$Xe and $^3$He are used. A co-magnetometer is often used which attempts to sample the same volume as that of the neutrons in order to reduce sensitivity to possible spatial variations of the magnetic field. As indicated above, the measured or calculated (for the case of $^3$He~\cite{FLAM07}) EDM of the magnetometer species must be much smaller than the intended sensitivity of the neutron EDM experiment. The false EDM due to the geometric phase can be minimized with sufficient control of magnetic field gradients. 

Most of the new round of neutron EDM experiments rely on trapped free neutrons, so called Ultra Cold Neutrons (see Ref.~\cite{UCNBOOK}), produced by a new generation of higher efficiency "superthermal" neutron moderators. These moderators use cold neutrons (produced from either a reactor or proton beam spallation) interacting with phonons in either superfluid $^4$He (LHE-II)~\cite{GP77} or solid deuterium (SD$_2$)~\cite{GB83}. Such sources may be able to increase the trapped UCN densities by factors of 100-1000 compared to the mechanical ``moderator" (a.k.a. turbine) used for the present best limit on the neutron EDM~\cite{NEDM15}. Another potential improvement could be increasing the maximum electric field since the EDM sensitivity is linear in this quantity. Previous experiments have been limited to about 10-12 keV/cm due to the materials needed to confine the neutrons and the magnetometers. One of the future experiments plans to use the superfluid $^4$He moderator as the UCN storage medium potentially allowing higher electric fields due to the high breakdown strength of liquid helium. This concept, first discussed in Ref.~\cite{GL94}, also allows polarized $^3$He, injected into the superfluid, to be used as a co-magnetometer and simultaneously as a monitor of the neutron's precession frequency using the highly spin-dependent neutron-$^3$He capture cross section which produces scintillation light in the LHe due to the energetic protons and tritons in the final state. 

As discussed above, measurements of the neutron EDM look for a change in the precession frequency of the neutron's magnetic dipole moment precessing in a uniform magnetic field associated with the application of a uniform electric field. Most experiments use the Ramsey separated-oscillatory-field technique~\cite{RAM49} in which the polarization of a sample of precessing, polarized UCN is measured after an observation time $T_m$. The measurement time is fixed for measurements with E-field parallel and anti-parallel to the magnetic field. Any change in the final-state polarization correlated with the direction of the E-field is a signature of an EDM. For the new concept~\cite{GL94} discussed above, two different EDM measurement techniques will be used. In the first, the E-field-dependent change in precession frequency is measured via the scintillation light oscillation frequency for the n-$^3$He capture, with the $^3$He oscillation frequency measured with SQUID magnetometors. For the second technique, a non-resonant AC magnetic field, perpendicular to the constant precession field, is used to modify both the neutron and $^3$He precession frequencies (so-called spin dressing~\cite{CH69}) such that they both have the same precession frequency in the absence of an applied E-field; called critical spin dressing~\cite{GL94}. In this case the application of an E-field will lead to a detectable change in the precession frequency of the neutron, with a negligible change in the $^3$He precession, as a signature of the neutron EDM.

Many of these concepts and technologies are incorporated in the next generation of experiments described below. 


\section{Present Status of Neutron EDM Searches}

\begin{figure}[htb]
\includegraphics[height=4.5in]{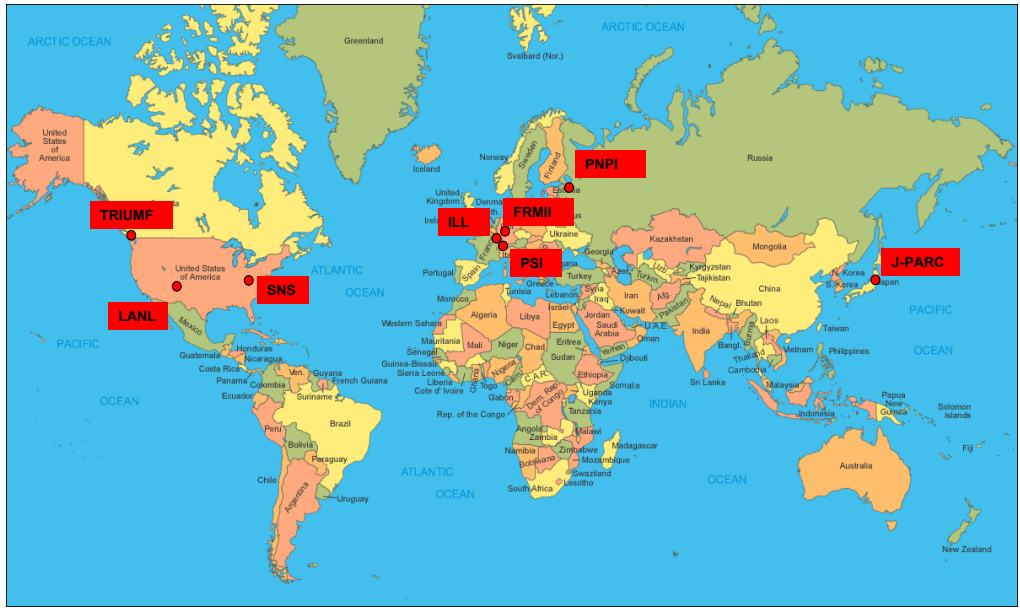}
\vskip -.7in
\caption{Facilities where new neutron EDM experiments with improved sensitivity are under development.}
\label{fig:map}
\end{figure}

Improving the sensitivity to the neutron EDM by one to two orders-of-magnitude is a technically challenging task. But due to the strong theoretical motivation to search for new sources of CP violation, a large number of new experiments are underway, taking advantage of many of the technological advances discussed in the previous section. Since these experiments require an intense source of free neutrons they are located at research facilities with either research nuclear reactors or proton-induced spallation accelerators. The location of these facilities, where new neutron EDM searches are being developed, is shown in Fig.~\ref{fig:map}.
\vskip -.3in

\newpage
Starting several years ago the community began organizing a series of workshops where the technical advances and challenges are presented and compared amongst the various experimental efforts. The most recent such workshop was held in October 2017 and the program of presentations for the workshop is available in Ref.~\cite{NEDM2017}. Most of these new experiments utilize one of the new types of superthermal UCN sources discussed above. A summary of these new higher sensitivity searches for the  neutron EDM using UCN are compared in Table~\ref{tab:nedm}, where the type of UCN source, the techniques used to measure the EDM and the expected sensitivity (at 90\% confidence level) are indicated. A new cold neutron beam concept is also being considered~\cite{NEDMBEAM} with development work underway. 

\begin{table}[t]
\begin{center}
\begin{tabular}{l|c|c|c|l}  
Experiment &  UCN Source &  Cell & Technique & Sensitivity  \\ 
& & & & x10$^{-28}$ e-cm\\ \hline
 ILL-PNPI  &   ILL turbine    &  Vac.     &    Ramsey technique for $\omega$ & Phase 1 $<$ 100\\ 
&  PNPI-SD$_2$ & &E=0 cell for magentometry & Phase 2 $<$ 10 \\ \hline
 PSI nEDM&  SD$_2$    &  Vac.      &    Ramsey, Cs + Hg co-mag. & Phase 1 $<$ 100\\ 
& & &$^3$He, Hg, Cs magnetom. & Phase 2 $<$ 20 \\ \hline
 Munich/ILL &  SD$_2$@FRMII  &  Vac.     &  Ramsey + Hg co-mag. +& Phase 1 $<$ 50 \\ 
& LHe@ILL& &external $^3$He/Cs mag. & Phase 2 $<$ 5\\ \hline
 TRIUMF& LHe-II  &    Vac.      & Ramsey technique &  $<$ 50  \\
(TUCAN)& & & with Hg + Xe co-mag.  & \\ \hline
 SNS nEDM &  LHe-II   &     LHe     &  Cryo-HV, &  $<$ 5  \\
& & &n-$^3$He capture for $\omega$, & \\ 
& & &SQUIDS+Critical dressing& \\ \hline
JPARC &  SD$_2$     &    Vac.    &  Under Devlopment &  $<$ 5(?) \\ \hline
 LANL EDM &  SD$_2$     &    Vac.     &  Ramsey with Hg&  $<$ 50 \\ \hline
\end{tabular}
\caption{Summary of the ongoing searches for a neutron EDM using Ultra Cold Neutrons. The second column refers to the type of superthermal UCN source, third colum refers to the medium used in measuring the EDM (Vac. = vacuum), the fourth column refers to the measurement technique and the last column is the potential EDM sensitivity after several years of running. }
\label{tab:nedm}
\end{center}
\end{table}

\begin{figure}[h]
\includegraphics[height=5.in]{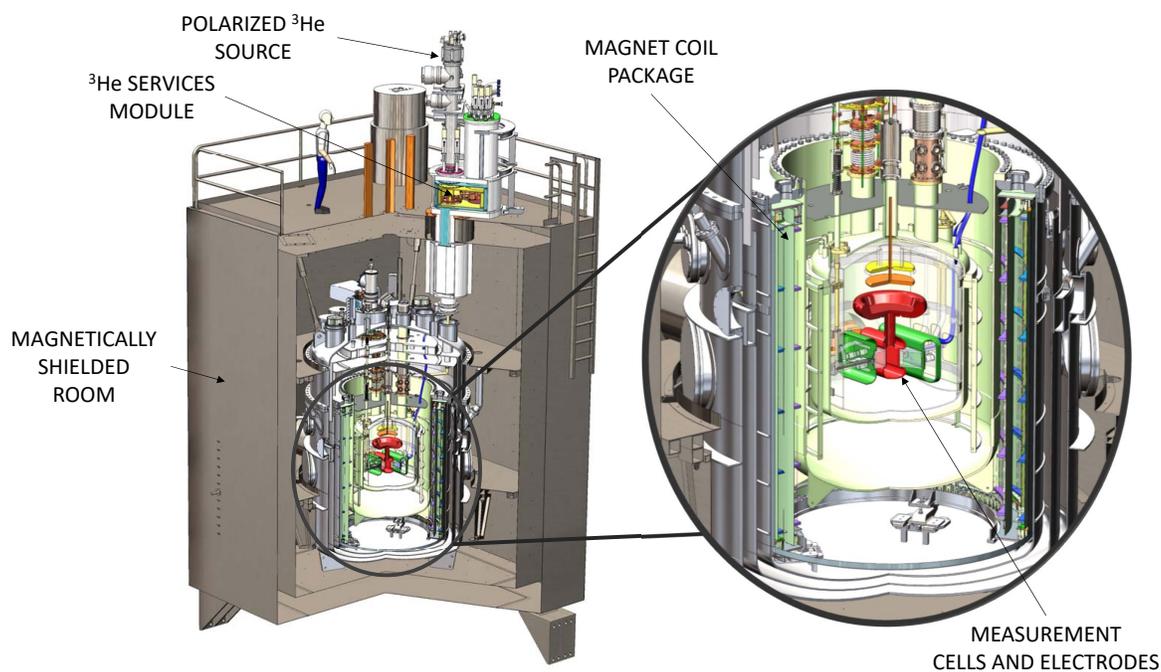}
\caption{Key components of the nEDM experiment at the Spallation Neutron Source at Oak Ridge National Laboratory.}
\label{fig:magnet}
\end{figure}

Several of these new UCN-based searches are already taking data in an initial phase of the experiment and most of them anticipate taking high sensitivity data within the next three years. The JPARC experiment has not finalized their experimental configuration, while the SNS experiment at the Oak Ridge National Laboratory requires five more years to complete construction and commissioning, but should begin data-taking with one of the highest sensitivities for these new experiments. 
This latter experiment is an all-cryogenic search wherein the UCN are produced in a pair of superfluid $^4$He filled measurement cells that are submerged in about 1500 liters of LHe. The measurement cells must be purged of $^3$He (down to a $^3$He to $^4$He ratio of $N_3/N_4 <10^{-12}$) prior to the injection of highly polarized $^3$He with $N_3/N_4 = 10^{-10}$. Due to the cryogenic nature of the experiment, the magnetic field system can use a nearly hermetic superconducting shield and persistent current for the main magnetic holding field.  A schematic of the experimental configuration is shown in Fig.~
\ref{fig:magnet}. A detailed discussion of the design of this new experiment is being prepared for publication~\cite{NEDMSNS}.

\newpage

\section{Summary}

The science reach of future neutron EDM experiments is discussed as well as the new technical advances that are being applied to extend the sensitivity of the experiments from one to two orders-of-magnitude. The technical methods, sensitivity and status of these new searches are discussed.

\end{document}